\documentstyle[twocolumn,aps,epsf]{revtex}

\begin{document}

\twocolumn[
\hsize\textwidth\columnwidth\hsize\csname@twocolumnfalse\endcsname
\draft

\title{Evidence for field-induced excitations in low-temperature thermal conductivity
of Bi$_2$Sr$_2$CaCu$_2$O$_8$}
\author{Herv\'e Aubin, Kamran Behnia}
\address{Laboratoire de Physique des Solides (CNRS), Universit\'e Paris-Sud, 91405 Orsay, France }
\author{Shuuichi Ooi, Tsuyoshi Tamegai}
\address{Department of Applied Physics, The University of
 Tokyo, Hongo, Bunkyo-ku, Tokyo, 113-8656, Japan}
\date{July 2, 1998}
\maketitle

\begin{abstract}
The thermal conductivity ,$\kappa$, of Bi$_2$Sr$_2$CaCu$_2$O$_8$ was studied as a function of magnetic field. 
Above 5 K, after an initial decrease, $\kappa(H)$ presents a kink followed by a plateau, as 
recently reported by Krishana et al..  By contrast, below 1K, the thermal conductivity 
was found to \emph{increase} with increasing field. This behavior is indicative of a finite 
density of states and is not compatible with the existence of a field-induced fully gapped 
$d_{x^{2}-y^{2}}+id_{xy}$ state which was recently proposed to describe the plateau regime. 
Our low-temperature results are in agreement with recent works predicting a field-induced 
enhancement of thermal conductivity by Doppler shift of quasi-particle spectrum. 

\end{abstract}

\pacs{74.25.Fy, 74.72.Bk, 72.15.Eb, Submitted to Phys. Rev. Lett.}
]

Until very recently, thermal conductivity in the vortex state of high-T$_c$ cuprates was analyzed
through a picture which was originally developped in the context of conventional superconductors. 
According to this scheme,  vortices constitute new scattering centers for heat carriers and their
introduction leads to a decrease in heat conductivity. 
Even in the absence of a satisfactory account of vortex-quasi-particle interaction in unconventional
superconductors, this picture provided a qualitative explanation of the observed field-induced 
decrease in $\kappa(H)$ in the investigated region of field-temperature plane . 
However,  recent results reported by the Princeton group on the thermal conductivity in the mixed 
state of Bi$_2$Sr$_2$CaCu$_2$O$_8$ (Bi2212)\cite{krishana97} and La$_{2-x}$Sr$_x$CuO$_4$\cite{ong97} 
have demonstrated the need for a more vigorous exploration of transport in the vortex state of the 
cuprates. Notably, their study of $\kappa(H)$ in Bi2212\cite{krishana97} indicated that
for temperatures below  20K, above a threshold field $H_{k}(T)$, $\kappa$ becomes insensitive
to the magnitude of magnetic field applied along the c-axis. The authors proposed that 
$H_{k}$---which presents a  rough $T^{2}$ dependence---is the signature of a phase transition 
and the high-field plateau  regime represents a new  quasi-particle-free superconducting state.
In their suggested scheme, the low-energy excitations of the parent $d_{x^{2}-y^{2}}$ order 
parameter is supressed at $H_{k}(T)$ due to the opening of a large $d_{x^{2}-y^{2}}+id_{xy}$ 
gap\cite{laughlin97} in the whole Fermi surface. 

In this letter we present a study of the field-dependence of thermal conductivity
at various temperatures down to 0.1K. Our results reveal an  increasing $\kappa(H)$ at 
subkelvin temperatures. We will argue that this is a strong indication for presence
of field-induced  delocalized  excitations  and
as such rules out the hypothetical fully-gapped state invoked\cite{krishana97,laughlin97} to explain
the plateau observed at higher temperatures. Moreover, our results highlight the relevance of 
field-induced Doppler shift of energy spectrum which---as recently pointed 
out\cite{Hirschfeld98}---has been hitherto neglected in analyses of thermal transport 
in the cuprates. 

Fig. 1 presents the thermal conductivity of two optimally-dopped Bi2212 crystals at temperatures 
just above 5K (T$_c$ is 89K for sample 1 and 88K for sample 2).  Our results in this temperature range confirm the behavior 
originally reported by Krishana et al.\cite{krishana97}: after an initial drop, 
$\kappa(H)$ presents  a kink and then becomes quasi-constant. These authors
attributed the  plateau of $\kappa(H)$ to a sudden disappearance of 
quasi-particle (QP) transport, presumably consequent to the opening of a gap induced 
by the magnetic field.
However, as we already argued in a comment presenting the data on sample 1\cite{Aubin98}, 
the existence of a strong hysteretic behavior weakens seriously this interpretation.
Indeed, if the high-field state was a superconducting state free of quasi-particles 
, one can not see why it should show such an extreme sensitivity to the details 
of the sample's thermo-magnetic history. In their reply\cite{krishana98}, 
Krishana et al. confirmed the existence of a hysteretic 
behavior with a sample-dependent intensity. They suggested that the disorder 
in the crystallographic structure could 
induce textures and defects in the vortex lattice forbiding the phase 
transition occuring at $H_{k}(T)$ to go to completion everywhere
in the crystal. In this way, $d_{x^{2}-y^{2}}$ domains with low-energy 
excitations would survive in the  gapped $d_{x^{2}-y^{2}}+id_{xy}$ 
superconducting state. They further argued that in the sweep-down 
trace of $\kappa(H)$, the flux distribution would be more uniform, with a 
lower density of  $d_{x^{2}-y^{2}}$ domains, and consequently, a lower
density of electronic heat carriers. However, this proposed explanation 
of hysteresis is unsatisfactory. It supposes that sweeping  the 
magnetic field up to $10H_k(T)$ leaves the density of 
the hypothetical domains unchanged---which is necessary 
to keep a so flat plateau. Moreover, it fails to explain the sign of the
hysteresis in sample 2.

To follow the temperature dependence of $H_{k}(T)$ we studied the thermal conductivity
of sample 2 at lower temperatures (T < 1K). For these measurements, instead of 
sweeping the magnetic field at a constant temperature, we used an alternative procedure: a 
constant magnetic field was applied above T$_c$ and then  thermal conductivity
was measured as a function of temperature. This was to avoid 
the well-known problems related to inhomogeneous penetration of vortices in a 
zero-field-cooled sample. Indeed, by 
studying the magnetization of the sample at T = 5K, we found the field of complete 
penetration to be 3500 Oe. For this reason, it is not
possible  to resolve a possible kink in  $\kappa(H)$ for fields smaller than 3500 Oe
which is the expected magnitude of H$_K$   below 4 K. On the other hand, by cooling the
sample below T$_c$ in a finite magnetic field,  one would expect to attain a homogeneous field 
in the sample even at small fields. 

%%%%%%%%%%%%%%%%%%%%%%%%%%%%%%%%%%%%%%%%%%%%%%%%%%%%%%%%%%%%%%%%%%%%%%%%%%
\begin{figure}
\epsfxsize=8.5cm
$$\epsffile{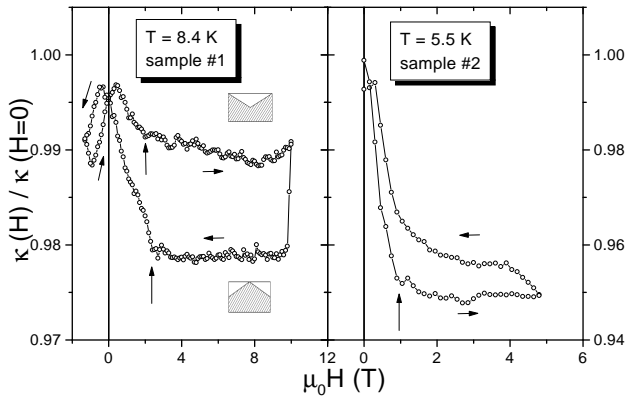}$$
\caption{ Field dependence of the thermal conductivity $\kappa(H)/\kappa(H=0)$ for
two different samples. Note the initial decrease, the kink, the ``plateau'' and
the large hysteresis.}
\label{fig1}
\end{figure}
%%%%%%%%%%%%%%%%%%%%%%%%%%%%%%%%%%%%%%%%%%%%%%%%%%%%%%%%%%%%%%%%%%%%%%%%%%%%
Fig. 2 shows the temperature
dependence of thermal conductivity $\kappa(T)/T$ for various magnitudes of
magnetic field. We found that even at the lowest temperature investigated 
(T $\simeq$ 0.1K) $\kappa(T)$ does not present a cubic term characteristic of 
phonon conductivity with constant mean-free-path. Thus, the separation of
the electronic and lattice components of zero-field thermal conductivity is not 
straightforward. Here we will concentrate on the effect of the magnetic field 
which can be analyzed indepedently.

As seen in the figure, at this temperture range, $\kappa$ is mostly enhanced 
when the sample is cooled down below T$_c$ in a finite magnetic field. Fig \ref{fig3}
shows the field profile of thermal conductivity extracted from $\kappa(T)$  data at various
fields together with $\kappa(H)$ at 5.5 K already  shown in Fig. 1 and obtained by 
sweeping upwards the magnetic field. The figure shows that at low temperatures,
instead of presenting a plateau, $\kappa(H)$ increases with the magnetic field. Furthermore, 
this field-induced enhancement of thermal conductivity becomes more pronounced with decreasing 
temperatures. This behavior has strong implications for the debate
on the origin of the plateau feature observed at higher temperatures. 

We begin by noting that lattice heat transport can only be reduced by magnetic 
field. The only way to explain a field-induced \emph{increase} in thermal conductivity 
is to invoke an enhancement of the available excitations of the superfluid condensate.
There are two possible origins for unpaired electrons in the mixed state. First, there are
localized excitations of the normal core. The second source is provided by the Doppler shift 
of QP energy spectrum due to the superfluid flow around the vortices.

%%%%%%%%%%%%%%%%%%%%%%%%%%%%%%%%%%%%%%%%%%%%%%%%%%%%%%%%%%%%%%%%%%%%%%%%%%%%
\begin{figure}
\epsfxsize=8.5cm
$$\epsffile{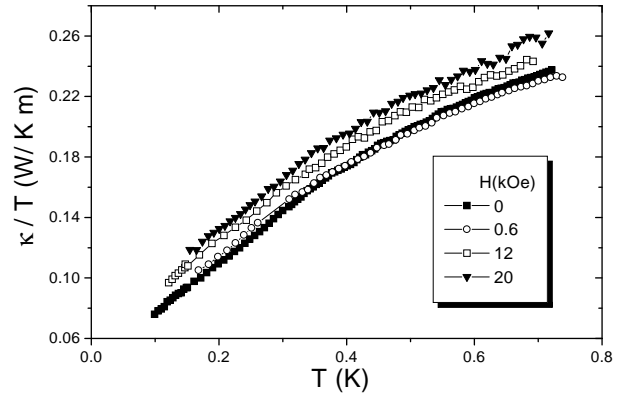}$$
\caption{Temperature dependence of the thermal conductivity $\kappa(T)/T$ with 
a magnetic field applied above T$_c$. Note the crossing of the 0.6 kOe and 0.0 kOe curves.} 
\label{fig2}
\end{figure}
%%%%%%%%%%%%%%%%%%%%%%%%%%%%%%%%%%%%%%%%%%%%%%%%%%%%%%%%%%%%%%%%%%%%%%%%%%%%
Vortex cores can  be excluded as a plausible source of heat transport in our context
of investigation. Let us recall that in a s-wave superconductor, the energy of the first bound 
state inside the vortex cores is $\Delta_{0}^{2}/2E_{F}$\cite{Caroli64}, which for cuprates would 
yield a fraction of meV. In a
$d_{x^{2}-y^{2}}$ type superconductor, according to analytical\cite{ichioka96} and 
numerical\cite{Tesanovic} works, one do not expect bound states
inside the vortex cores at all. Nevertheless, tunneling
spectroscopy measurements have shown the existence of bound states inside the
vortex cores of YBCO\cite{aprile95}, but the energy of the
first one is 5.5 meV (60K). In Bi2212, the very recent detection of vortices by 
tunneling spectroscopy \cite{Renner} revealed a pseudo-gap behavior in the vortex core and no
evidence for bound states. Considering all these points, it is fairly sure that at temperatures
as low as 0.18K, the electronic states inside the vortex cores---if they exist---would not be
occupied. Furthermore, they would be localized and could not contribute to 
heat conduction. 

Extended excitations associated with supercurrents around vortices constitute the only 
other possible source of field-induced excess conductivity. As first pointed 
out by Volovik\cite{Volovik93}, for an anisotropic superconducting gap, their contribution 
will dominate the variation of the density of states, even at small magnetic fields. 
They are induced by the Doppler shift of energy spectrum due to a finite local value of
superfluid velocity $\overrightarrow{v_{S}}$. This effect is
negligible in conventional superconductors at low magnetic fields, when the
isotropic gap $\Delta_{0}>>\mathbf{p}\overrightarrow{v_{S}}$. For a  $d_{x^{2}-y^{2}}$ gap
, on the other hand, the density of these excitations increases as $\sqrt{H}$, and simply
reflects the linear variation of the density of states with the energy, 
i.e. $N(\varepsilon)/N_{0}\sim\varepsilon/\Delta$. The basic relationship 
can be derived easily : the energy of the quasi-particles being shifted from :
$\varepsilon\rightarrow\varepsilon-\mathbf{p}\overrightarrow{v_{s}}(r)$, then
the increase of the density of states calculated on a vortex lattice cell will be
: $N(H)\sim\frac{1}{R^{2}}\int_{0}^{R}v_{s}(r)rdr$ where $R\sim1/\sqrt{H}$.
Because $v_{s}(r)\sim1/r$, we find that $N(H)\sim1/R\sim\sqrt{H}$. Experimental evidence
for the existence of this type of field-induced delocalized excitations in high-T$_c$
cuprates was first provided by specific heat measurements on YBCO\cite{Moler94}. 
This field-induced shift of the energy spectrum produces effects similar to an increase in 
temperature. This equivalency is at the origin of the scaling behavior predicted
for the field-temperature dependence of specific heat in unconventional 
superconductors\cite{Lee97,Volovik97}. The
experimental observation of these scaling relations in specific heat of YBCO\cite{Revaz98}and 
thermal conductivity of UPt$_3$\cite{Suderow98}, indicates that a convincing picture of the effects of 
the magnetic field on the density of states in unconventional superconductors is now emerging.
Recently, K{\"u}bert and Hirschfeld\cite{Hirschfeld98} observed that one implication of this picture for the transport
properties of the mixed state of the cuprates is to expect the thermal conductivity to increase
with the application of a magnetic field at low temperatures in opposition to the decrease 
experimentally reported at intermediate temperatures. Our results constitute
the experimental confirmation of such a field-induced increase in $\kappa$ by 
lowering temperature\cite{taillefer}. They can be naturally explained if one 
assumes the presence of nodes (or very deep minima) in the superconducting gap 
of Bi2212 at this range of temperature and field.

In spite of a large background due to lattice contribution to heat conduction, we can proceed
to a more quantitative check of this picture. The inset of Fig. \ref{fig3} shows the variation of 
$\kappa(H)/\kappa(0)$ versus $\sqrt{H}$ at T=180 mK. As expected, the increase in $\kappa(H)$ is
proportional to $\sqrt{H}$. However, at these low temperatures, 
a reliable comparison with the theory would necessarily include the effects of the impurity 
band states\cite{Hirschfeld98b}. Indeed, the magnitude and the exact field dependence
of the thermal conductivity depends strongly on the energy dependence of the density
of states $N(\varepsilon)$ which is in turn a function of concentration of impurities and the 
phase shift introduced by their scattering potential. 

One remarkable feature of recent theoretical works on heat transport by field-induced
quasi-particles close to the nodes is the prediction of a non-monotonic behavior of
$\kappa$ at finite temperature\cite{Barash1998,Hirschfeld98}. This is a consequence of the 
energy-dependence of the relaxation time of the QP scattered by impurities. In presence of
magnetic field, the Doppler shift of excitations due to the finite local superfluid velocity 
leads to a decrease in the relaxation time at low fields which can exceed the parallel rise in 
$N(\varepsilon)$ and disrupt the monotonic increase of  thermal conductivity. Hence, this
theory is able to provide a unique explanation both for a monotonically increasing $\kappa(H)$
at low temperatures and a non-monotonical behavior at higher temperatures without
invoking vortex scattering of quasi-particles. In this picture, the position of the minimum
in $\kappa(H)$ is predicted to be $H/H_{c_2}\simeq(k_{B}T/a\Delta_0)^{2}$ where $H_{c_2}$
is the upper critical field, $\Delta_0$ is the gap maximum over the Fermi surface, and $a$ is a
vortex-lattice-dependent constant of order unity. 
A close examination of Fig. \ref{fig2} shows that  $\kappa(T)$  curves at H = 600Oe  and
H =0 intersect at T= 0.4 K which is indicative of a
temperature-dependent minimum in $\kappa(H)$. This becomes evident in the right panel of
 Fig.\ref{fig3}, where a logarithmic plot highlights the low-field region. Assuming 
H$_{c2}$= 200 T, a=0.5 and  $\Delta_0=2.14k_BT_c$, a rough agreement is found 
between the theoretically expected position of the minimum and the experimental data for 
T=0.55K. It is important to emphasize that
the ordinary electron-vortex scattering which dominates the transport
properties of the mixed state in conventional superconductors---which should be present
to some extent in the cuprates---is neglected in this model.

%%%%%%%%%%%%%%%%%%%%%%%%%%%%%%%%%%%%%%%%%%%%%%%%%%%%%%%%%%%%%%%%%%%%%%%%%%%%
\begin{figure}
\epsfxsize=8.5cm
$$\epsffile{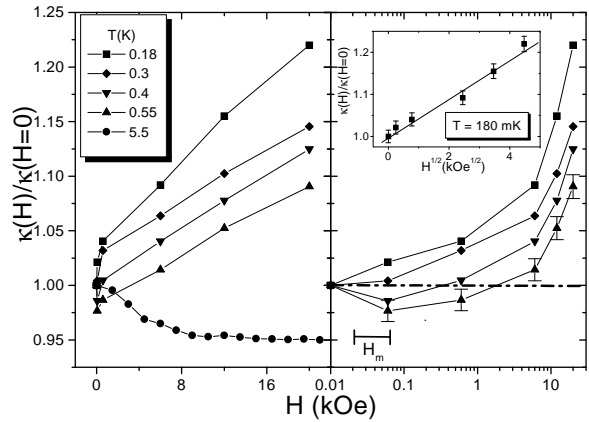}$$
\caption{Field profile $\kappa(H)/\kappa(H=0)$ for various temperatures.
 Left panel : linear-linear. Right panel : linear-log.
At low temperatures, thermal conductivity increases with the magnitude of 
the magnetic field. H$_m$ represents the expected position of minimum for 
0.55K(see text).The inset in the right panel
shows $\kappa(H)/\kappa(0)$ versus $\sqrt{H}$ at 0.18 K.}
\label{fig3}
\end{figure}
%%%%%%%%%%%%%%%%%%%%%%%%%%%%%%%%%%%%%%%%%%%%%%%%%%%%%%%%%%%%%%%%%%%%%%%%%%%%%

What are the implications of our results for the present debate on the nature of
the anomaly discovered by Krishana et al.? The grey zone in  Fig. 4 represents
our zone of exploration in  the (H,T) plane where the thermal conductivity 
was found to increase with the magnetic field  due to a finite density
of states. By comparing the position of this region
with the extrpolation of the H$_k$(T) line to low temperatures, we can consider two 
possibilities: i)  $H_{k}(T)$, experimentally established from 5 to 20 K, instead of 
extrapolating to low temperatures, increases above the grey region as shown by the dot line.
This is improbable. ii) $H_{k}(T)$ extrapolates to low temperatures with roughly the same
slope(the solid line was theoretically predicted by Laughlin\cite{laughlin97}),but this 
transition can not be seen in the thermal conductivity profile simply due to the lack
of resolution. In this case, the basic augmentation of the thermal
conductivity with a so low magnetic field imply that this new high field state
must have a non-zero residual density of states, and consequently deep minima (or nodes) in the
superconducting gap.

%%%%%%%%%%%%%%%%%%%%%%%%%%%%%%%%%%%%%%%%%%%%%%%%%%%%%%%%%%%%%%%%%%%%%%%%%%%%%
\begin{figure}
\epsfxsize=8.5cm
$$\epsffile{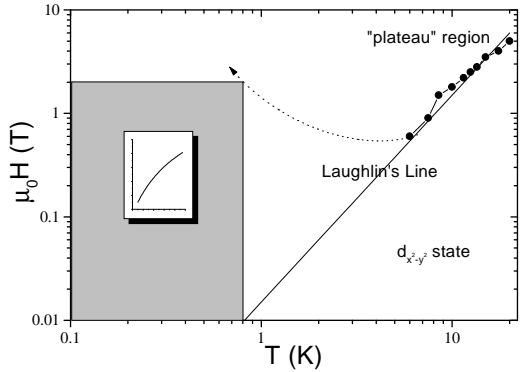}$$
\caption{$\kappa$ was found to increase with magnetic field in the grey zone of 
the (H,T) plane. Full circles represent experimental  H$_k$(T) extracted from 
Ref. 1. The solid line is the theoretical prediction by Laughlin.}
\label{fig4}
\end{figure}
%%%%%%%%%%%%%%%%%%%%%%%%%%%%%%%%%%%%%%%%%%%%%%%%%%%%%%%%%%%%%%%%%%%%%%%%%%%%%

These observations do not rule out a phase
transition. Indeed, the kink observed is robust from one sample to
another, and is not obviously related to a phase transition in the
vortex lattice. However, if this phase transition in the electronic
system exists, its nature or consequences on the electronic heat carriers are 
different from what is simply  expected for a  state with a large uniform gap suppressing the
excitations of the superconducting state in the whole sample. Already, when presenting
his model on field-induced transition to a $d_{x^{2}-y^{2}}+id_{xy}$\cite{laughlin97} state, 
Laughlin pointed out that the expected gap is too small to account for a complete
vanishing of QP transport. Moreover, he stressed  the inherent difficulty in explaining a 
large field-induced gap exceeding $k_BT_c$, the temperature at which the transition occurs. 

At this stage, a broken time-reversal symmetry associated with the splitting of 
$d_{x^{2}-y^{2}}$ state remains the most serious track to explain the origin of the strong 
hysteretic behavior. Indeed, the field profile in the sample is associated with a superfluid current of the order of J$_c$,
the critical current. When reversing the magnetic field ramping direction, the most
dramatic change in the sample is the direction of the field profile and the way of
circulation of the screenings currents. The sensitivity of thermal conductivity to
the direction of circulation of these screenings currents may
indicate that these situations are not symmetric, which would be the case in presence
of a current associated with  broken time-reversal symmetry in the superconducting state. Some other experimental 
results have been recently interpreted as signatures of a splitting 
of the $d_{x^{2}-y^{2}}$ state.  For example, Franz and Tesanovic\cite{Tesanovic} have proposed that the
existence of a $d_{x^{2}-y^{2}}+id_{xy}$ state could be at the origin of the bound states observed in
the vortex cores of  YBCO by tunneling spectroscopy\cite{aprile95}. Furthermore, the  observation of a
spontaneous split of the zero-bias conductivity peak at 7K in YBCO\cite{covington97}, was attributed to a phase
transition at the sample surface from a pure $d$-wave to a 
time-reversal violating state.

In conclusion, our results show that the new high-field superconducting state in Bi2212 identified
by the emergence of a plateau in the field-dependence of the thermal conductivity 
cannot be simply attributed to a vanishing of low-energy excitations in 
the superconducting state. At low-temperature, $\kappa$ increases with a weak magnetic
field, implying  a non-zero density of states. 

We are indebted to L. Fruchter for magnetization measurements. Fruitful discussions
with him and with L. Taillefer, P. Hirschfeld and Y. Barash are acknowledged. This 
work was partly supported by Grant-in-Aid for Scientific Research from the Ministry of
Education, Science, Sports and Culture, Japan.


\begin{thebibliography}{10}
\bibitem{taillefer}
A similar bahvior has been recently observed in YBCO (L. Taillefer, Private communication).
\end{thebibliography}
\end{document}